\documentstyle[12pt]{article}

\begin{document}
\hoffset 0.5cm
\voffset -0.4cm
\evensidemargin 0.0in
\oddsidemargin 0.0in
\topmargin -0.0in
\textwidth 6.7in
\textheight 8.7in

\begin{titlepage}

\begin{flushright}
PUPT-1653\\
hep-th/9610125\\
October 1996
\end{flushright}

\vskip 0.2truecm

\begin{center}
{\large {\bf Probing bound states of D-branes$^*$}}
\end{center}

\vskip 0.4cm

\begin{center}
{Gilad Lifschytz}
\vskip 0.2cm
{\it Department of Physics,
     Joseph Henry Laboratories,\\
     Princeton University, \\
     Princeton, NJ 08544, USA.\\ 
    e-mail: Gilad@puhep1.princeton.edu }

\end{center}

\vskip 0.8cm

\noindent {\bf Abstract} 
A zero-brane is used to probe non-threshold BPS
bound states of ($p$, $p+2$,$p+4$)-branes.
At long distances the stringy calculation agrees with the supergravity
calculations.
The supergravity description is given, using the interpretation
of the $D=8$ dyonic membrane as the bound state of a two-brane 
inside a four-brane.  
We investigate the short distance structure
of these bound states, compute the phase shift of the scattered zero-brane
and find the bound states  characteristic size.
It is found
that  there should be a supersymmetric
solution of type IIa supergravity, describing a bound state
of a zero-brane and  two orthogonal two-brane,  
all inside a four-brane , with
an additional unbound zero-brane. We comment on the relationship
between $p$-branes and $(p-2)$-branes.

\noindent                
\vskip 0.8 cm

\begin{flushleft}

$^*$ {\small This work was supported in part
by the National Science Foundation grant PHY-9315811.}

\end{flushleft}

\end{titlepage}

\section{Introduction}
D-branes \cite{dlp,gr}
 have emerged as important objects in string theory. It is thus
important to understand various properties of these objects.
Properties of bound states at threshold are important
tests for string duality conjectures \cite{wit1,sen}.
In this paper we explore, non-threshold BPS
\footnote{All these states should form an algebra, which
may elucidate some underlying structure of string theory \cite{hamo}.}
bound states between D-branes of dimension
$p$, $p+2$ and $ p+4$. 
This is done by probing them with another D-brane.
In particular
we investigate the short distance structure of these bound states.

D-branes
can be used to probe sub-string distances
\footnote{See \cite{stpr} for a treatment of 
strings scattered off D-branes.}. 
There is 
a correspondence between the infra-red world-volume 
theory and the ``space-time'' description of the D-brane moving 
in a background \cite{dbac}. At sub-string scales, one does not expect
to have a space-time description. In fact, one generally does not know
an appropriate description of the physics. However, 
when D-branes are involved
one can readily see that the short distance physics is governed
by the light modes of the open-superstring \cite{dkps} 
(or by the light modes of the closed string together with an infinite
tower of massive modes), so a space time description is not
available, but the physics is under control. 
In cases where some of the supersymmetries are not
 broken another description 
comes in, namely that of a moduli space of a 
supersymmetric world-volume theory. 
If one-quarter of the supersymmetries
are un broken, the metric on the moduli space is protected from 
higher loop corrections. 
In effect, a space-time description means that low-velocity particles
follow geodesics of some metric, but this is exactly what a moduli
space means, so even at sub-string scales there are situations
where the physics of two objects (or more) can be effectively 
described by a space-time \cite{dkps}. However the crucial dependence 
of this description on the probing agent means that it is not
a universal description, and thus might differ from our notion of
space-time.
Classically, the D-brane are singular at $r=0$. The
world-volume description of the region near the D-brane shows
that as another D-brane approaches the ``singularity'', the physics
is described by a transition 
to another branch, and is not singular.

The string description of the bound states we are going to consider
is given by open-superstring
ending on the brane with some world-volume gauge fields turn on. 
As 
discussed in \cite{tow,cjp} this endows a $p$-brane with 
an $RR$ charge of a 
$(p-2)$-brane. Further, these are BPS states \cite{gutgre},
thus this string description,
 is a reasonable description for these bound-states.
 We  compare this
description to a supergravity description at long distances and 
find that they agree.
 
After preliminaries in section $(2)$, the bound state of a two-brane
and a zero-brane $(2-0)$ is studied in section (3). We compare the 
string description to a supergravity description by comparing the 
long-range potential between a zero-brane and the $(2-0)$ bound state.
In section (4) we treat the $(4-2)$ bound state. We compute the
long-range, velocity dependent potential between the bound state
and a zero-brane and compare them to a super-gravity calculation.
We also compute the phase shift the scattered zero-brane acquires
after scattering from the bound-state at short distances. From the phase 
shift
we compute the absorption probability of 
the zero-brane by this bound state
and the size of the bound-state. 
Section (5) is devoted to the study of the $(4-2-2-0)$
bound state, where the two two-branes are orthogonally embedded in the
four-brane. We end with conclusions.


\section{Probing the ($p$, $p-2$) bound state}

Starting with the  action of a $p$ brane moving in a background, 
let us concentrate on the coupling to
the RR sector \cite{doug}
\begin{equation}
I_{RR}=T_{p}\int_{W_{p+1}}Str \ \  C \wedge e^{2\pi \alpha' F}.
\label{birr}
\end{equation}
Here $T_{p}=\sqrt{\pi}(4\pi^{2} \alpha ')^{(3-p)/2}$ \cite{pol,cjp},
$F=dV-\frac{B}{2\pi\alpha '}$, $V$ is
the world volume gauge field and $B$ is the two-form NS-NS gauge field.
A $p$ brane with a constant magnetic field of the world volume 
gauge field strength
$F$ will carry a $RR$ charge depending on the form of $F$.
Here we are assuming that the $p$ brane is compactified on a torus.
One can think of this
as representing the bound state of a $p$ brane with various
 lower dimensional branes. We will compare our result to 
the supergravity description of some of these objects.
A constant magnetic
 field on the $p$
brane is relatively easy to treat and we will mainly scatter zero-branes
 off  various configuration in order 
to learn about the property of the bound states.
This will be done by computing the one loop vacuum amplitude for open 
superstring with the appropriate boundary conditions \cite{pol}.

The one loop vacuum amplitude is the phase shift of the
probe after the scattering \cite{bac}, and defines a potential $V(r^{2})$
through the equation
\begin{equation}
A(b,v)=-\int d\tau V(r^{2}=b^{2}+\tau^{2}\frac{v^2}{1-v^{2}}),
\label{defpot}
\end{equation} 
where $b$ is the impact parameter and $v$ is the velocity of the 
D-brane probe.
The short distance
behavior is governed by the light open string modes while the
long distance is governed by the light closed string modes. Thus when
approximating the integrals one should take care to match those two 
approximation \cite{dkps}.

Given two identical parallel D-brane with the same condensation $F$ on
their world-volume, it is known that the one loop vacuum amplitude
is just the same as when the condensation is zero, except for
 a multiplicative 
factor of $(1+2\pi \alpha' f^2)$ \cite{acny,bacpor,gutgre}
 ($f$ is the non zero entry of $F$). 
This factor expresses the change in the mass of the brane, due
to the binding with a lower brane, and the change in the $RR$ charge. 
Thus the mass of the bound state of a $p$-brane and a $(p-2)$-brane
is 
\begin{equation}
m^{2}_{(p, p-2)}=m^{2}_{p}+m^{2}_{p-2},
\end{equation} 
which is the dual to the mass formula in \cite{wit}.
The 
brane's original charge is un-modified, but there is an additional charge 
density of
order $f$ on each brane, which is the $RR$ charge of the lower dimensional
brane. Some D-brane configurations, with world volume gauge 
field turn on, were considered in \cite{gutgre,wit,ck,li,bf,bdl,as}.

For completnese we give the following.
In terms of $q=e^{-\pi t}$, we define
\begin{eqnarray}
f_{1}(q) & = & q^{1/12} \prod_{n=1} (1-q^{2n}). \\
f_{2}(q) & = & \sqrt{2} q^{1/12} \prod_{n=1} (1+q^{2n}). \\
f_{3}(q) & = & q^{-1/24} \prod_{n=1} (1+q^{2n-1}). \\
f_{4}(q) & = & q^{-1/24} \prod_{n=1} (1-q^{2n-1}).
\end{eqnarray}

The limit of $t \rightarrow 0$ one has,
\begin{eqnarray}
f_{1}(q) & \rightarrow & \frac{1}{\sqrt{t}}e^{-\pi/(12t)}. \\
f_{2}(q) & \rightarrow & e^{\pi/(24t)}(1-e^{-\pi/t}). \\
f_{3}(q) & \rightarrow & e^{\pi/(24t)}(1+e^{-\pi/t}). \\
f_{4}(q) & \rightarrow & \sqrt{2}e^{-\pi/(12t)}.
\end{eqnarray}

We will also find it convenient to have the behavior of
the $\Theta(\nu t,it)$ (Jacobi theta functions)
 in the limit $t \rightarrow 0$
\begin{eqnarray}
\frac{\Theta_{1}(i\epsilon t,it)}{\Theta^{'}_{1}(0,it)} & \rightarrow &
-e^{(\pi \epsilon^{2} t)}\frac{t}{i}\frac{\sin (\pi \epsilon)}{\pi}. \\
\frac{\Theta_{2}(i\epsilon t,it)}{\Theta_{2}(0,it)} & \rightarrow &
e^{(\pi \epsilon^{2} t)}(1+4\sin^{2}(\pi \epsilon)e^{-\pi/t}). \\
\frac{\Theta_{3}(i\epsilon t,it)}{\Theta_{3}(0,it)} & \rightarrow &
e^{(\pi \epsilon^{2} t)}(1-4\sin^{2}(\pi \epsilon)e^{-\pi/t}). \\
\frac{\Theta_{4}(i\epsilon t,it)}{\Theta_{4}(0,it)} & \rightarrow &
e^{(\pi \epsilon^{2} t)}\cos(\pi \epsilon).
\end{eqnarray}


\subsection{Compact Branes}
When some of the space times coordinate are compact their
 effect on the configuration of
the D-branes depends on whether the compact dimensions are
 an $NN$, $ND$ or $DD$
directions. In the case a $NN$ direction is compact the
 integral over the momentum
in that direction becomes a sum over the allowed momenta.
 If the compact
direction is  a $DD$ direction there is no momentum integral 
to begin with, however
there are infinite number of  open string configuration 
that wrap around
the compact direction. Thus the mass of the open string is now
$M^2 =\frac{b^2}{(2\pi\alpha')^{2}} + \frac{1}{\alpha '}
\sum (oscillators) + (nR/\alpha')^2$, 
 $R$ is the radius of the compact direction,
and there is a  string
configuration for each $n$.
In the case of a $ND$ 
direction there are no momentum 
integrals and no winding modes, 
thus there is no change in the 
one loop computation. This is of course what one 
expects from T-duality which
changes a $DD$ direction to a $NN$ direction but 
the number of $ND$ directions
remain the same.

The one loop amplitude for a configuration of a $p$ brane 
and an $l$ brane moving  parallel
to each other with one $NN$ direction compactified is ($L=2\pi R$)
\begin{equation}
A=\frac{C_{l-1}}{2\pi}\int \frac{dt}{t} e^{-(\frac{b^2 t}{2\pi \alpha'})} 
(8\pi^2 \alpha' t)^{-(\sharp NN -1)/2} 
\Theta_{3}(0,8i\pi^{2} \alpha ' t/L^2)
 B \times J.
\label{comnn}
\end{equation}

$B$ and $J$ are the usual contribution from the bosonic and 
fermionic oscillators
respectively.
Similarly for a compactified $DD$ direction one finds 
\begin{equation}
A=\frac{C_{l}}{2\pi}\int \frac{dt}{t} e^{-(\frac{b^2 t}{2\pi \alpha'})} 
(8\pi^2 \alpha' t)^{-(\sharp NN)/2} \Theta_{3}(0,itL^2/2\pi^{2}\alpha ')
 B \times J.
\label{comdd}
\end{equation}
For example, in the case of $p=6, l=2$ and one of the $NN$ directions being
compact, one can calculate 
the $v^2$ term of the potential
 to be ($\beta=\frac{r^2 L^2}{16\pi^{2} \alpha'^{2}}$ )
\begin{equation}
V=-\frac{\pi C_1 v^2 L^2}{(8\pi^2 \alpha')^{2}}
 \frac{\coth(\sqrt{\beta})}{\sqrt{\beta}},
\end{equation}
from which the moduli space metric can be read off.
When $\beta$ is large the potential falls like 
$r^{-1}$ as expected from a six-brane,
and when $\beta$ is small the metric falls like $r^{-2}$ 
as expected from a five brane. 

From equation (\ref{comnn}) the important scale that
determines the behavior of a system with $NN$ compact directions,
is $bL$. If $bL$ is large than the
system will behave as if it is un-compact and vise versa.
So if we probe deeply a system compactified on an $NN$ direction
then it will behave as if the compactification scale is small. Similarly
If $L$ is small but we go far away the system will behave as if it is 
un-compactified. For a compact $DD$ direction  the relevant
scale is of course $b/L$. 

In the next sections we will have to deal with compact dimensions that are 
different than $NN, DD$ or $ND$. 
We will be faced with compact coordinates
that satisfy a $D$ or $N$ boundary condition on one end of the string
and some condensation on the other end, we shall call them
$NF$ and $DF$ conditions. When an $NF$ or $DF$ direction is compact 
things are different. For a $DF$ condition there will not be any
momenta integral but will be something 
like a winding, and vise versa for the
$NF$ coordinates. In order to avoid this complication we will always assume
that the radius of compactification is large enough as to neglect those 
effects, even in the large $r$ limit,
 and we will treat those directions as if they are un-compactified
(from the modes point of view). 

\section{(2-0) bound state}
For the two-brane there is only one relevant term in the expansion 
(\ref{birr}) and it is 
$A \wedge F$, where $A$ is the $RR$ gauge field carried by the zero-brane.
We will assume that the two-brane is compactified on $T^{2}$.
If one chooses 

\[
F=\left(
\begin{array}{ccc}
0 & 0 & 0 \\
0 & 0 & f \\
0 & -f & 0
\end{array}
\right)
\]
then,
\begin{equation}
\int A \wedge F \rightarrow \int f d^{2}\sigma \int A d \tau.
\end{equation}
Requiring $\int f =2\pi$ gives the two-brane action a term
$T_{0}\int Ad\tau$, which is the coupling of a zero-brane to a $RR$
background.

As $f=$const, the zero-brane $RR$ charge of this configuration
 is proportional to 
$fL^2$ where $L^2$ is the area of the compactified two-brane.

Let us  compute the velocity-dependent potential, between the (2-0) bound 
state and another zero-brane moving with velocity $v$.
The one loop amplitude (the phase shift) takes the form 
($\tan(\pi \epsilon) = 2\pi\alpha' f$, $\tanh(\pi \nu)= v$),
\begin{equation}
A=\frac{1}{2 \pi}\int \frac{dt}{t} e^{-(\frac{b^2 t}{2\pi \alpha'})} B \times J,
\label{a20}
\end{equation}
\begin{eqnarray}
B& = &\frac{1}{2}f_{1}^{-6}\Theta^{-1}_{4}(i\epsilon t)
\frac{\Theta_{1}' (0)}{\Theta_{1}(\nu t)}, \\
J& = &\{ -f_{2}^{6}
\frac{\Theta_{2} (\nu t)}{\Theta_{2}(0)}
\Theta_{3}(i\epsilon t, it)+f_{3}^{6}\Theta_{2}(i\epsilon t, it)
\frac{\Theta_{3} (\nu t)}{\Theta_{3}(0)} \nonumber\\ 
 & + & if_{4}^{6} 
\frac{\Theta_{4} (\nu t)}{\Theta_{4}(0)}\Theta_{1}(i\epsilon t)\}.
\label{bj20}
\end{eqnarray}

The existence of the $NS(-1)^F$ sector, the third
term in equation (\ref{bj20}), is a 
consequence of the new boundary condition
for the open super-string. 
Instead of having 2 $ND$ coordinates which gives fermionic
zero modes in that sector those two coordinates now have
 different boundary conditions $FD$ \cite{acny},

\begin{eqnarray}
\partial_{\sigma}X^{\mu}+2\pi \alpha' F^{\mu}_{\nu}\partial_{\tau}X^{\nu}
& = & 0 \ \ \ (\sigma=0), \\
\partial_{\tau} X^{\mu} & = & 0 \ \ \ (\sigma=\pi). 
\end{eqnarray}
Similarly for the fermionic coordinates \cite{bacpor}.

We treat the case where the non-zero component of $F$ are constant.
One can solve for the modes and compute the one loop amplitude.
A short cut to the right answer is to start with the expression
in \cite{bacpor} which is for the case of an electric field on both 
branes (in their case a $9$-brane).
 To get a magnetic field one just substitutes 
$\epsilon \rightarrow i\epsilon$ and to get a Dirichlet boundary condition
on one of the branes one can formally take the condensation on that brane
to $\infty$. This has the following effect, one substitutes
\begin{eqnarray}
\Theta_{1}(i\epsilon) & \rightarrow & i\Theta_{4}(i\epsilon). \\
\Theta_{2}(i\epsilon) & \rightarrow & \Theta_{3}(i\epsilon). \\
\Theta_{3}(i\epsilon) & \rightarrow & \Theta_{2}(i\epsilon). \\
\Theta_{4}(i\epsilon) & \rightarrow & i\Theta_{1}(i\epsilon).   
\end{eqnarray}
Further more, because of the Dirichlet boundary condition on 
one end there are 
no zero-modes in the bosonic sector. The velocity dependence is
as in \cite{bac}, thus we end up with equations (\ref{a20}-\ref{bj20}).

Of course, when $\epsilon$ goes to zero in equations (\ref{a20}-\ref{bj20})
one gets back 
just the expression for a zero-brane scattered off a two-brane.

Let us check that one gets the right charge for the zero-brane inside the
two-brane. Taking only the $RR$ sector (the third term in equation 
(\ref{bj20})) one finds that
the charge per unit volume of the zero-brane inside the two-brane 
is
proportional to $\sim T_{2}\tan(\pi \epsilon)=2\pi T_{2} \alpha 'f$ 
exactly as expected, and the $RR$ 
sector has the right sign to represent 
interaction of two zero-brane of same charge.

One can compute the one loop amplitude in various limits.
When the distance between the branes $r$ is large 
one gets for the velocity dependent potential
\begin{equation}
V=-\Gamma(5/2)\frac{(2+ 2\sin^{2}(\pi \epsilon) +2\sinh^{2}(\pi \nu) -
4\sin(\pi \epsilon)\cosh(\pi\nu))}
{\cos(\pi \epsilon)\sqrt(8\pi^{2} \alpha ')}
(\frac{2\pi \alpha '}{r^{2}})^{5/2}.
\end{equation}
Where $\cosh(\pi\nu)=\frac{1}{\sqrt{(1-v^{2})}}$ and 
$\sinh(\pi\nu)=\frac{v}{\sqrt{(1-v^{2})}}$.

These results of course hold with some modification to all
T-dual configurations. For instance the long range potential
between a bound state of a four-brane and a two-brane  and another
two-brane parallel to the one inside the four-brane is, ($Q_{4}$
is the four-brane charge)
\begin{equation}
V_{string} \sim -Q_{4}\frac{(2+ 2\sin^{2}(\pi \epsilon) 
+2\sinh^{2}(\pi \nu) -
4\sin(\pi \epsilon)\cosh(\pi\nu))}{\cos(\pi \epsilon)}
r^{-3}.
\label{422st}
\end{equation}

We can compare this result,
 with the conjectured supergravity description of the bound state
of a two-brane inside the four-brane \cite{ilpt}. 
The supergravity configuration of a two-brane inside a four-brane,
was first derived as the $D=8$ dyonic membrane.
It will be convenient
to write down its eleven-dimensional interpretation. The metric takes the 
form
\begin{eqnarray}
ds_{11}^{2} & = & (H \tilde{H})^{1/3}[H^{-1}
(-dt^{2}+dy_{1}^{2} +dy_{2}^{2})
+\tilde{H} ^{-1}(dy_{3}^{2}+dy_{4}^{2}+dy_{5}^{2}) \nonumber \\
 & + & dx_{1}^{2}+ \cdots dx_{5}^{2}]. \nonumber \\
F_{4}^{(11)} & = &\frac{1}{2}\cos (\zeta) \star dH +
\frac{1}{2}\sin (\zeta)  dH^{-1} \wedge dt \wedge dy_{1} 
\wedge dy_{2} \nonumber\\
 & + & \frac{3 \sin (2\zeta)}{2\tilde{H} ^{2}}dH\wedge dy_{3} \wedge dy_{4}
\wedge dy_{5}. 
\label{sg42}
\end{eqnarray}
Here $H=1+\frac{\gamma}{r^{3}}$, 
 $\tilde{H}=1+\frac{\gamma \cos^{2}(\zeta)}{r^{3}}$ and $\star$ is the
Hodge dual in $R^{5}(x_{1} \cdots x_{5})$.
Now this is the metric of the eleven dimensional two-brane inside a 
five-brane, but when one considers any of the $y_{i}$ $i=3-5$ as the
eleventh direction we get a two-brane inside a four-brane.  Further the
two-brane charge is $Q_{2} \sim \gamma \sin (\zeta)$ and the
four-brane charge $Q_{4} \sim \gamma \cos (\zeta)$.
If we choose $y_{5}$ as the eleven dimension (so its radius is small), the
other two $y$'s are also compactified but on a large circle, as 
discussed in section (2). 

Now one can calculate the velocity 
dependent potential between a two-brane
and this bound state, where the two-branes are parallel,
 using the metric and gauge fields in 
equation (\ref{sg42}). This is easiest done in the static gauge, and
 one can readily use the formulas in \cite{dklt} to find,
\begin{equation}
V_{sugra} \sim \frac{\gamma}{r^{3}}
[4sin(\zeta) +2\cos^{2}(\zeta)-4-v^{2}\cos^{2}(\zeta)]
\label{422sug}
\end{equation}
Comparing this to equation (\ref{422st}) we find they do not agree.
The reason for that is that while in the supergravity calculation
 we have worked in the ``static gauge'' in which the time like parameter
of the world volume is equal to $X_{0}$, this is not the case in
the string calculation. These expressions then identify potentials
in two different reference frames. The string calculation can 
easily be converted to this frame. 
Observe that in the string calculation
we have taken the expression for the potential of the form
\begin{equation}
A=-\int d \tau V(r^{2}=b^{2} +\tau^{2} \sinh^{2}(\pi \nu)),
\label{av}
\end{equation}  
so that $\tau \neq X_{0}$. In order to get the string theory answer
corresponding to the observer $\tau=X_{0}$ one just needs to multiply
equation (\ref{422st}) by a factor 
$\frac{v}{\sinh(\pi\nu)}=\cosh^{-1}(\pi\nu)$. Then the two expressions,
that of the string theory and that of the supergravity agree to order
$v^{2}$ when one identifies $\zeta=\pi\epsilon$.

When $r$ becomes very small the appropriate expansion is of 
$t \rightarrow \infty$. We introduce a 
cutoff $\Lambda$, and the $t$ region
$0 \rightarrow \Lambda$ is governed by the light closed string modes
which make a non-singular contribution of order $\sim r^2$. Then 
one gets (for $v=0$)
\begin{equation}
A \approx \int_{\Lambda}^{\infty}
 \frac{dt}{t} e^{-t(\frac{b^2 }{2\pi \alpha'}-\pi/2 +\pi\epsilon)} 
(8\pi^2 \alpha' t)^{-(1/2)}
\end{equation}
Now when $b <\pi/2 -\pi\epsilon$ a tachyon develops in the open string 
spectrum and the expression becomes complex \cite{bansus}. 
However this happens
at a slightly smaller distance than in the pure two-brane case.

What happens when $\epsilon$ grows, the largest it 
could be is $\epsilon=1/2$.
Huresticly as $\epsilon$ grows one gets more and more ``towards''
a Dirichlet boundary condition on the two-brane, 
and the tachyonic instability
starts at  smaller distances.
One can see that the constant
term and the $v^{2}$ term in the potential goes to zero while the 
$v^{4}$ term grows. 

Notice that a bound state configuration of $(2-0)$ is dual
to a bound state of a D-string and an elementary string.
In the first
case this is described by turning on a magnetic field on
the world-volume and in the second it is by turning on
an electric field in the world volume \cite{wit}.

\section{(4-2) bound state}
The world-volume of the four brane is five dimensional and the
 coupling we are
going
to consider in this section is $C \wedge F$, where C is the $RR$
 three form gauge
field coupled to the two-brane. The four-brane will be wrapped around
$T^{2}$. 
Taking 

\[
F=\left(
\begin{array}{ccccc}
0 & 0 & 0 & 0 & 0 \\
0 & 0 & f & 0 & 0 \\
0 & -f & 0 & 0 & 0 \\
0 & 0 & 0 & 0 & 0 \\
0 & 0 & 0 & 0 & 0
\end{array}
\right)
\]

the $F\wedge F$ coupling will not contribute, 
so we will have a bound state of a two-brane
 and a four-brane. Of course one can orient the two-brane inside the four
brane in more than one way, by choosing which elements of $F$ will be
non zero. The membrane world volume occupies the directions where
$F=0$, so if $F_{12} \neq 0$ then the membrane is in the $3,4$ direction
inside the four-brane, and it is the $1,2$ direction which is compact
(not the directions the two-brane occupies).

The phase shift for a moving zero-brane probe (with velocity $v$) 
in the presence of the $(4-2)$ bound state
is ($\tanh(\pi \nu) =v$, $\tan(\pi \epsilon)=2\pi \alpha' f$),
\begin{equation}
A=\frac{1}{2\pi}\int \frac{dt}{t} e^{-(\frac{b^2 t}
{2\pi \alpha'})} B \times J.
\end{equation}
\begin{eqnarray}
B & = & \frac{1}{2}f_{1}^{-4}f_{4}^{-2}\Theta^{-1}_{4}(i\epsilon t)
\frac{\Theta_{1}' (0)}{\Theta_{1}(\nu t)}. \\
J & = & \{ -f_{2}^{4}f_{3}^{2}
\frac{\Theta_{2} (\nu t)}{\Theta_{2}(0)}
\Theta_{3}(i\epsilon t, it)+f_{3}^{4}f_{2}^{2}\Theta_{2}(i\epsilon t, it)
\frac{\Theta_{3} (\nu t)}{\Theta_{3}(0)}\}.
\end{eqnarray}

As $\epsilon \rightarrow 0$ one gets the result for a zero-brane
 and a pure four-brane.
This expression can be evaluated at various limits.
 Let as first compare the string
description with the supergravity description. 
The appropriate limit is then 
$r \rightarrow \infty$, so $t \rightarrow 0$, which is the range 
when the mass-less
closed string modes dominate.
One finds a velocity dependent potential,
\begin{equation}
V=-\Gamma(3/2)\frac{(\sin^{2}(\pi \epsilon) 
+\sinh^{2}(\pi\nu))}{\cos(\pi \epsilon)\sqrt(8\pi^{2} \alpha ')}
(\frac{2\pi \alpha '}{r^{2}})^{3/2}.
\label{420st}
\end{equation}

We turn now to the supergravity description.
If there is a zero-brane interacting with the $(4-2)$ bound state,
 one can compute \cite{gil} the
velocity dependent potential from null 
geodesics on the  metric (\ref{sg42}).
Then one finds
\begin{equation}
V_{sugra} \sim \frac{-Q_{4}}{r^{3}}\frac{(\frac{v^{2}}{1-v^2}+\sin^{2}(\zeta))}
{\cos(\zeta)}
\label{sgpot}
\end{equation}

Equation (\ref{sgpot}) agrees with equation (\ref{420st})  
when one identifies $\zeta=\pi\epsilon$
\footnote{Notice that we do not have the problem of converting 
the string calculation to the static gauge, because the 
supergravity calculation is done differently than in the previous section}.
In the string
description  $\epsilon=1/2$ describes an infinite condensation of 
two-branes on the four-brane. On the supergravity side this is the case
$\zeta=\pi/2$, which makes the supergravity solution into a pure two-brane.
A condensation of infinitely many two-branes on the four-brane
have turned it into a two-brane.

Let us now turn to the case that $r$ is shorter than the string scale.
As $v \rightarrow 0$ the potential between the zero brane and the $(4-2)$
bound state is, 

\begin{equation}
V=-\frac{\Gamma(-1/2)}{\sqrt{8\pi^2 \alpha '}}
[(\frac{r^2}{2\pi \alpha '}-\pi \epsilon)^{1/2}
+ (\frac{r^2}{2\pi \alpha '}+\pi \epsilon)^{1/2}- 2(\frac{r^2}
{2\pi \alpha '})^{1/2}]
\end{equation} 

The first two-terms are from the $NS$ sector of the open 
string and the last from the $R$ sector. It exhibits the characteristic
of a stretched string potential between the branes.
A feature that is due to the probing agent, the zero-brane.
The difference in the ground state energy of the different sectors,
translates into different effective length for the stretched string.
Now if one views $\epsilon$ as a parameter that can change, then
this expression tells as that $\epsilon$  will want to grow.
This means that if we have fixed the charges on the brane, then it is
the volume of the compactified brane that will tend to decrease.  

To order $(\pi\epsilon)^{2}$
The potential becomes
\begin{equation}
V=-\frac{\sqrt{\pi}(\pi \epsilon)^{2}}{2\sqrt{8\pi^2 \alpha '}}
(\frac{2\pi \alpha '}{r^{2}})^{3/2}
\end{equation} 
So
for small $r$ and large $r$ the potential agree to order
 $\epsilon^2$, which will
 enable us later to approximate some integrals in a simple way. 
This agreement is a 
residue of the 
supersymmetry present when $\epsilon=0$.

At non-zero velocity one can compute the phase shift of the 
scattered zero-brane.
For small $r$ we can take the $t$ 
integral limits from $0 \rightarrow \infty$
because in this case the $t \rightarrow 0$ limit of the open string
is the same as the closed string.

We consider the case where the velocity ($v$) is small. Then
one finds ($\pi \nu \approx v$)
\begin{equation}
A=\int \frac{dt}{t} e^{-(\frac{b^2 t}{2\pi \alpha'})}(\tan(\frac{vt}{2})+
\frac{\cosh(\pi \epsilon t)
-1}{\sin(vt)})
\label{a42s}
\end{equation}
This gives,
\begin{equation}
e^{iA}= \frac{\Gamma[\frac{ib^2}{4v\pi \alpha '}+
\frac{1}{2}-\frac{i\pi\epsilon}{2v}]
\Gamma[\frac{ib^2}{4v\pi \alpha '}+\frac{1}{2}+\frac{i\pi\epsilon}{2v}]}
{\Gamma[\frac{ib^2}{4v\pi \alpha '}+1]\Gamma[\frac{ib^2}{4v\pi \alpha '}]}.
\label{ps42}
\end{equation}

Equation (\ref{a42s}) exhibits a tachyonic instability at 
$b^2 < 2\pi^{2} \alpha ' \epsilon$, this will translate after analytic 
continuation to a large  imaginary part of equation (\ref{ps42}),
which would mean a very small norm for the scattered wave function
(i.e absorption).
An incoming zero-brane in a plane wave state with velocity in the $z$ 
direction, will be
 multiplied by the phase shift (equation \ref{ps42})
after scattering, so
\begin{equation}
e^{ikz} \rightarrow e^{ikz +iA(b^{2}=x^{2}_{\perp},v)}.
\end{equation}

The norm of the outgoing wave function is given by, 
\begin{equation}
|e^{iA}|=[\frac{\sinh^{2}(\frac{b^2}{4v \alpha '})}{\cosh^{2}
(\frac{b^2}{4v \alpha '})+
\sinh^{2}(\frac{\pi^2 \epsilon}{2v})}]^{1/2}.
\label{norm}
\end{equation}
If the norm of the wave function is much less than $1$, it signals the
breakdown of the WKB-Eikonal approximation. For low $v$ this will
happen when $b^2 < 2\pi^{2} \alpha ' \epsilon$.

If we Fourier transform the outgoing wave we will get the scattering 
amplitude as a function of the incoming momenta ($k=\frac{v}{g}$)
\begin{equation}
f(k,\theta) \sim \exp{[-\sqrt{2} k \sin(\theta/2)
(\sqrt{\pi\epsilon +(gk)^2}-
\pi\epsilon)^{1/2}]}.
\label{sca42}
\end{equation}
In the limit $v \gg \epsilon$ we get 
\begin{equation}
f(k,\theta) \sim e^{-\sqrt{2}\sin(\theta/2)(kl_{p}^{11})^{3/2}}
\end{equation} 
as in \cite{dkps}, which shows that the 
physical scale is the eleven dimensional
Planck length  
$l_{p}^{11}=g^{1/3}l_{s}$ \cite{kp}.
In the limit $\epsilon \gg v$ one finds
\begin{equation}
f(k,\theta) \sim e^{-\sin(\theta/2) \frac{g}{\sqrt{\pi\epsilon}} k^{2}}
\end{equation} 
However in this limit it is easy to see that the approximation 
is not valid.

Now $(1-|e^{iA}|^{2})$ is the probability of absorbing a zero-brane
by this bound state at impact parameter $b$. At very low velocities
($v \ll \epsilon$) one sees that the probability is $\sim 1$ for 
$b/2\pi \alpha ' <\pi \epsilon$ and zero otherwise.
One can interpret a scale $r_{0}$ at which there is a 
large absorption probability, as
giving the effective scale of the bound state. Of course
this depends on the probing agent. thus the above result
 is what
we expect from a state with characteristic length scale
in string units of $\sqrt{\pi\epsilon}$. For very small $\pi \epsilon$
this gives a characteristic scale for the bound state 
(as seen by the zero-brane) 
$\sim \frac{(2 \pi \alpha ')}{L}$.

The probability of absorbing a zero-brane by this bound state when the 
zero-brane is an incoming state $e^{ikz}\phi(b)$ is
\begin{equation}
P_{abs}=\int d^{4} b |\phi(b)|^{2}(1-\|e^{iA}\|^{2})
\end{equation}
 If one assumes that $\pi^{2} \epsilon \ll v$ and that $\phi(b)$
is zero outside a region of volume $V_{4}$ and constant in it,
then one finds
\begin{equation}
P_{abs}\approx \frac{\Omega_{3}}{2V_{4}}[(4v\alpha')^{2} \ln 2+
(\pi \epsilon)^{2}(2\pi \alpha')^{2}(\frac{2}{3}\ln 2-\frac{1}{6})]
\label{abs}
\end{equation}
Where $\Omega_{3}$ is the area of the unit three-sphere.
The first term in (\ref{abs}) is present for the pure four-brane, where
it is interpreted as a signature for resonances \cite{dfs}. The
second term represents the effective size of the bound state.


\section{Probing the (4-2-2-0) bound state}
We take the four-brane to be wrapped around $T^{4}$. If one chooses

\[
F=\left(
\begin{array}{ccccc}
0 & 0 & 0 & 0 & 0 \\
0 & 0 & f & 0 & 0 \\
0 & -f & 0 & 0 & 0 \\
0 & 0 & 0 & 0 & f \\
0 & 0 & 0 & -f & 0
\end{array}
\right)
\]
then the four-brane will be endowed with two-brane charge ( two orthogonal 
two-branes) and
 zero-brane charge, due to the coupling
\begin{equation}
\frac{1}{2}A\wedge F \wedge F + C \wedge F.
\end{equation}
The mass of this bound state is
\begin{equation}
m^{2}_{4-2-2-0}=m^{2}_{4} +m^{2}_{2}+m^{2}_{2}+m^{2}_{0}
\end{equation}

The phase shift of a moving zero-brane in the
 background of this
bound state is 

\begin{equation}
A=\frac{1}{2\pi}\int \frac{dt}{t} e^{-(\frac{b^2 t}{2\pi \alpha'})}
 B \times J.
\label{amp4220}
\end{equation}
\begin{eqnarray}
B& = &\frac{1}{2}f_{1}^{-4}\Theta^{-2}_{4}(i\epsilon t)
\frac{\Theta_{1}' (0)}{\Theta_{1}(\nu t)}. \\
J& = &\{ -f_{2}^{4}
\frac{\Theta_{2} (\nu t)}{\Theta_{2}(0)}
\Theta_{3}^{2}(i\epsilon t, it)+f_{3}^{4}\Theta_{2}^{2}(i\epsilon t, it)
\frac{\Theta_{3} (\nu t)}{\Theta_{3}(0)} \nonumber \\
 & + & f_{4}^{4} 
\frac{\Theta_{4} (\nu t)}{\Theta_{4}(0)}\Theta_{1}^{2}(i\epsilon t)\}.
\label{bj4220}
\end{eqnarray}

When $v=0$ one finds that the one loop amplitude vanishes, due
 to one of the identities of the Jacobi functions  \cite{witwat}.
This means that a configuration
of a  (4-2-2-0) bound state of this type together with an additional 
zero-brane is a stationary solution of the supergravity equations, and
should  preserves  some super-symmetries, probably a quarter
\footnote{After writing this work, I have learned that a T-dual 
description of this configuration was discussed in \cite{bdl}}.

In the limit of large $r$ one finds a potential
\begin{equation}
V=-\Gamma(3/2)\frac{2(1-\cosh (\pi\nu))\sin^{2}(\pi \epsilon)+
\sinh^{2}(\pi\nu)}{\cos^{2}(\pi \epsilon)
\sqrt{(8\pi^{2} \alpha ')}}(\frac{2\pi \alpha '}{r^{2}})^{3/2}.
\end{equation}
Notice that to order $v^{2}$ one gets the same result as in the case 
of a zero-brane 
scattered off a four-brane \cite{gil}.

In the limit of small $r$ one finds
\begin{equation}
A=\int \frac{dt}{t} e^{-(\frac{b^2 t}{2\pi \alpha'})} \tan(vt/2).
\end{equation}
Exactly like in the case of the zero-brane and the pure four-brane.
One can evaluate the phase shift as in \cite{dkps}, 
\begin{equation}
e^{iA}= \frac{\Gamma[\frac{ib^2}{4v\pi \alpha '}+\frac{1}{2}]
\Gamma[\frac{ib^2}{4v\pi \alpha '}+\frac{1}{2}]}
{\Gamma[\frac{ib^2}{4v\pi \alpha '}+1]\Gamma[\frac{ib^2}{4v\pi \alpha '}]}.
\end{equation}

Thus the moduli space
is the same as in the case of a zero-brane moving in the background of 
a four-brane,
\begin{equation}
ds^{2}=\frac{1}{g}(1+\frac{g}{2r^{3}})(dr^{2}+d\Omega^{2}_{4}).
\end{equation}

Notice however that although the results are similar to those in the 
case where there is a zero-brane scattered off a pure four-brane, the
physics is different. For instance in the latter case at large distances
the physics is governed by gravity alone, while in the $(4-2-2-0)$ case
there are gauge field interactions as well.

If we assume that the condensation on the four brane was not the same
in both direction that is

\[
F=\left(
\begin{array}{ccccc}
0 & 0 & 0 & 0 & 0 \\
0 & 0 & f_1 & 0 & 0 \\
0 & -f_1 & 0 & 0 & 0 \\
0 & 0 & 0 & 0 & f_2 \\
0 & 0 & 0 & -f_2 & 0
\end{array}
\right)
\]

and $f_1 \neq f_2$, then the one loop has the same form as 
equation(\ref{amp4220}-\ref{bj4220}), with the change,
 $\Theta_{j}^{2}(i\epsilon t) \rightarrow 
\Theta_{j}(i\epsilon_1 t)\Theta_{j}(i\epsilon_2 t)$.
The configuration of this bound state with a zero-brane is not
static any more. Assuming we take a configuration of a four-brane 
two two-brane and a one-brane the
different $f$'s represent different areas for the two $T^2$'s.
 At short distances this gives a potential

\begin{equation}
V=-\frac{\Gamma(-1/2)}{\sqrt{8\pi^2 \alpha '}}
[(\frac{b^2}{2\pi \alpha '}-\pi (\epsilon_1-\epsilon_2))^{1/2}
+ (\frac{b^2}{2\pi \alpha '}+\pi (\epsilon_1 - \epsilon_2))^{1/2}
- 2(\frac{b^2}{2\pi \alpha '})^{1/2}]
\end{equation} 
As in the previous section $(\epsilon_{1}-\epsilon_{2})$ will tend to grow.
In this case this means that there is a force that will make the area
of one of the
$T^2$ larger than the other. So the space time will tend to 
develop very different scale in some directions.

The phase shift for the scattering of the zero-brane can be computed
for small $r$ and it has the form of equation (\ref{ps42}) with
the substitution $\epsilon \rightarrow (\epsilon_1-\epsilon_2) $.
This is also true for equations (\ref{sca42}-\ref{abs}).
The characteristic scale of the bound state
is $r^{2}_{0}\sim 2\pi^{2} \alpha' (\epsilon_1 -\epsilon_2)$. 

\section{Conclusions}

In this paper we give a string description of the bound states of
$(p, p+2, p+4) $ D-branes.
We compare the string description with a supergravity description,
using the interpretation of the dyonic membrane as the bound state of
a two-brane inside a four-brane. Both description agree at large
distances. 
We compute the velocity
dependent potential between a zero-brane probe and these bound states and
the phase shift of a scattered  zero-brane at short distances.
The size of the bound state as seen by a zero-brane is  estimated by
looking at the absorption cross section.
We find that the size of the bound state ($r_{0}$) is related
to the scale of the compact dimension ($L$) of the higher brane,
as $r_0 \sim L^{-1}$. 
The largest it could be is one half the string length.
In a certain range of parameters one finds that the high energy scattering
at fixed angle is governed by the eleventh dimensional Planck scale.

We have found that a special $(4-2-2-0)$ bound state does not exert a force
on an additional zero-brane. 
This is  
evidence that there should exist a solution of the
supergravity equations that corresponds to a (4-2-2-0) bound state
and another zero-brane, that preserves a quarter of the super-symmetries.
It will be interesting to find the corresponding super-gravity solution.  
The moduli space, in this case, turns out to be 
the same as in the pure
four-brane zero-brane case .This is also true for the long range 
interaction between this bound state and a 
zero-brane, even though the physics
of the pure four-brane zero-brane system looks  very different.

If one takes the limit of infinite condensation on one of the branes, we
saw it transformed that brane into a $p-2$-brane, this was seen in the
string and in the supergravity descriptions. This may suggest that
a $p$-brane is made out of infinitely many ($p-2$)-branes \cite{tow,bfss}.

\centerline{\bf{Acknowledgments}}
I would like to thank S. Deser, S. Mathur and S. Ramgoolam
for many helpful
discussions.

\end{document}